\documentclass[a4paper,11pt]{article}

\usepackage[titlenumbered,ruled]{algorithm2e}
\usepackage{fourier}
\usepackage{color}
\usepackage{graphicx}
\usepackage{url}
\usepackage[affil-it]{authblk}
\usepackage{amsmath}
\usepackage{wrapfig}
\usepackage{float}

\usepackage[T1]{fontenc}
\usepackage{times}

\usepackage{booktabs,caption}
\usepackage[flushleft]{threeparttable}
\usepackage{enumitem}
\usepackage{subfigure}
\DeclareUnicodeCharacter{2212}{-}
\newcommand{\comment}[1]{}

\newcommand{\rowname}[1]

\usepackage{mwe} 
\usepackage{multirow}
\usepackage{comment}
\usepackage{booktabs}
\usepackage{microtype}
\usepackage{tabularx}
\usepackage{authblk}

\begin{document}
\title{
An Ensemble Deep Learning Approach for COVID-19
Severity Prediction Using Chest CT Scans }
\author[1]{Sidra Aleem\thanks{Contributed equally}}
\author[1]{Mayug Maniparambil\thanks{Contributed equally}}
\author[2]{Suzanne Little}
\author[2]{Noel O\textquotesingle{}Connor}
\author[2]{Kevin McGuinness}
\affil[1,2]{ML Labs, SFI Centre for Research Training, Dublin City University}
\affil[ ]{\textit {\{sidra.aleem2, suzanne.little, noel.oconnor, kevin.mcguinness\}@dcu.ie}}
\affil[ ]{\textit {mayug.maniprambil@dcu.ie}}
\renewcommand\Authands{ and }

\date{}
\maketitle
\thispagestyle{empty}
\begin{abstract}
 
Chest X-rays have been widely used for COVID-19 screening; however, 3D computed tomography (CT) is a more effective modality. We present our findings on COVID-19 severity prediction from chest CT scans using the STOIC dataset. We developed an ensemble deep learning based model that incorporates multiple neural networks to improve predictions. To address data imbalance, we used slicing functions and data augmentation.
 We further improved performance using test time data augmentation. Our approach which employs a simple yet effective ensemble of deep learning-based models with strong test time augmentations, achieved results comparable to more complex methods and secured the fourth position in the STOIC2021 COVID-19 AI Challenge. Our code is available on online: 
 at: https://github.com/aleemsidra/stoic2021-baseline-finalphase-main.
\end{abstract}

\section{Introduction}
COVID-19 has created a global health crisis with millions of cases and deaths reported worldwide. Timely and accurate COVID severity prediction is essential for effective clinical management and treatment. 
Deep learning has shown tremendous potential in the medical domain. 
Automating the severity prediction of COVID-19 based on deep learning can lead to improved clinical workflow, resulting in faster diagnosis and better prognosis for severe COVID-19 cases.
 While a large number of studies have utilized deep learning methods for COVID-19 prediction based on chest X-ray data, CT scans have been found to be more effective in detecting COVID-19 positivity and severity \cite{aswathy2021covid}. Previous studies have primarily focused on COVID-19 positivity prediction or improving feature extraction methods. However, limited work has been done on COVID-19 severity prediction using CT scans, which are valuable for assessing lung condition, predicting COVID-19 severity, and detecting complications.

\section{Material and Method}\label{sec:method}
\begin{figure}
    \centering
    \includegraphics[width=0.9\textwidth, height = 5cm]{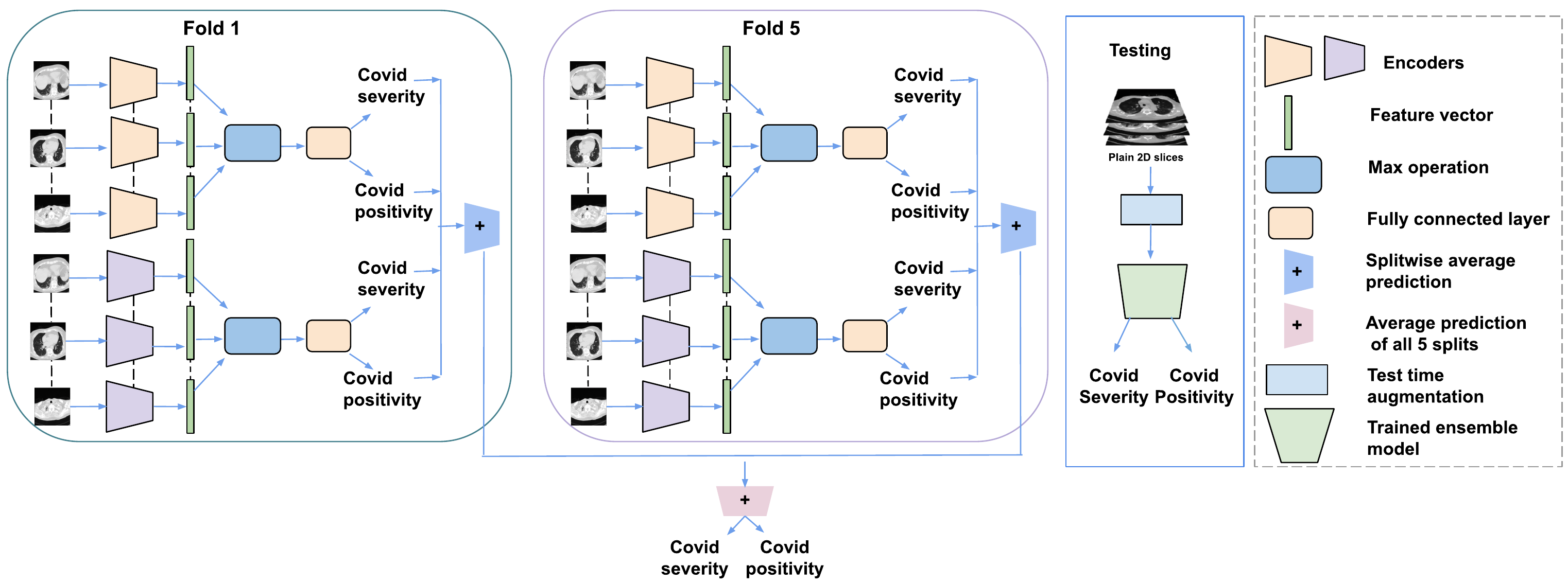}
    \caption{The schematic overview of our method.}
    \label{fig:arch}
\end{figure}

We employed an ensemble deep learning approach to predict COVID-19 severity in high-resolution 3D CT scans provided by the STOIC2021 COVID-19 AI Challenge. The scans had a spatial resolution of 512$\times$512 and a depth of 128 to 600 slices. To standardize the number of slices, we used a uniform sampling function, which samples 32 uniformly spaced slices from the slice dimension.  The radiodensity is measured using Hounsfield Units (HU), ranging from -1024 HU to 3071 HU, and stored as 12-bit numbers \cite{glide2013changes}.
However, directly scaling these values to $[0 - 1]$ results in low contrast images that make identification of COVID-19-related features difficult \cite{stern1995}.  To improve contrast, we utilize the lung window with a
window width of 1500 HU and a window level of -600 HU \cite{stern1995}.  
To deal with data scarcity, we partitioned the training set into five random splits. ResNet18 and MobileNetV3 are used for prediction. 
Figure \ref{fig:arch}. shows the overview of our method. For every split, each pre-processed slice $S_{i}$ is passed through both the ResNet18 and MobileNetV3 encoder $E$ to obtain feature vectors $z_{i} = f( S_{i})$. The maximally activated features are selected using the PyTorch $amax$ function along the slice dimension on the feature vectors of the slices as $z_{i}^{\max} = \/amax \left \{ z_{i} \right \}_{i=1}^{32}$ where $z_{i} \in R^{D}$. Only $z_{i}^{\max}$ feature vector is then used by the fully connected layer to form a prediction for COVID severity $\widehat{x_{s}}$ and COVID positivity $\widehat{x_{p}}$. The two models are combined through an ensemble technique, where the probabilities from both models are averaged to obtain the final prediction. Finally, the probabilities obtained from all five splits to obtain  are averaged together to get final prediction $\widehat{y_{s}}$ and $\widehat{y_{p}}$.

The experiments were carried out on publicly available STOIC data set \cite{stoic}.  It has CT scans of 2000 subjects and released under CC BY-NC 4.0 license. All the scans are in MHA format and have the size of 512x512. It has 301 severe cases. The performance was evaluated on STOIC private test set which consist of 7735 subjects. We faced challenges with limited STOIC public data (only 2000 subjects) and a heavily imbalanced data set (only 301 severe scans). To prevent bias towards the majority class, we used a balanced sampler that assigned weights to samples based on the inverse occurrence of their respective classes in the training set. Additionally, to avoid overfitting, we applied various image level augmentations. These augmentations led to improvements in our results (Table. \ref{tab:aug}), and we further enhanced them with test time data augmentation (TTA).

\section{Results and Discussion}
The STOIC data set has 10,735 subjects. It has been divided into a public set of 2,000 subjects, a qualification test set of 1,000 subjects, and a private training  set of 7000+ test subjects. The ground truth consists of two labels: COVID-19 positivity and COVID-19 severity. We experimented with multiple algorithms on the public set and chose the one that performed best on the public test set. We then used this algorithm on the qualification leaderboard (LB) test set. Based on the results obtained from the qualification LB, we were able to evaluate  the generalization performance of the model and submitted the improved version to final LB.
\begin{table}[!ht]
\centering
\caption{Performance evaluation based on (AUC Severity, AUC COVID).}
\begin{tabular}{lcccc}
\toprule
\multirow{2}{*}{\textbf{Model}} &\multicolumn{2}{|c|}{\textbf{STOIC Public Data}} & \multicolumn{2}{c}{\textbf{Qualification Leader board}} \\
 & \multicolumn{1}{|c|}{\textbf{AUC Severity}} & \multicolumn{1}{c|}{\textbf{AUC COVID}} & \multicolumn{1}{c|}{\textbf{AUC Severity}} & \multicolumn{1}{c}{\textbf{AUC COVID}}\\
 \midrule
CNN& 0.687  &   0.780 & -  &  - \\
ConvNext & 0.845 &  0.748 &   0.748 &  0.800\\
ResNet18 & 0.775 & 0.784 &  \textbf{0.752 } & 0.784\\
MobileNet & 0.817 & 0.780 & \textbf{0.779} & 0.735 \\
\bottomrule
\end{tabular}
\label{stoic_arch comparison}
\end{table}
Table~\ref{stoic_arch comparison}. shows the performance of various models on the public data set. ConvNext Tiny performed well on the public data set; however, when tested on the qualification LB, AUC severity dropped significantly to 0.748, indicating poor generalization performance. 

Alternative models were  also investigated: 3D CNN, MobileNetV3 small, and the use of encoders as fixed feature extractors along with age and sex metadata. These did not improve the predictive performance. Building on our initial results from Table \ref{stoic_arch comparison}, we conducted further experiments with ResNet18 and MobileNetV3. To evaluate the 

impact of augmentations on the overall predictions, we experimented with four set of augmentations. 1) Default: consisting of horizontal flip, random crop to 224x224, random gamma and color jitter with brightness 0.5, contrast 0.5 and  saturation 0.4. 2) Default + Strong: consisting of  safe rotate with limit 30, median blur, and the default augmentations. 3) Default + Strong + Mixup \cite{zhang2017mixup}: consisting of set 1, set 2 and mixup with $\alpha = 0.8$. 4) Default + Strong + Mixup + TTA: consisting of TTA, center crop, crops around four corners, safe rotate along -5, 10, +5 and all three sets as shown in Table \ref{tab:aug}.

\begin{table}[!ht]
\centering
\caption{Impact of augmentation on AUC Severity}
\begin{tabularx}{\linewidth}{l *{4}{>{\centering\arraybackslash}X}}
\toprule
\multirow{2}{*}{\textbf{Augmentation}} &\multicolumn{2}{c}{\textbf{Public data}} & \multicolumn{2}{c}{\textbf{Qualification LB}} \\
 & ResNet18 & MobileNetV3 & ResNet18 & MobileNetV3\\
 \midrule
Default& 0.775  &   0.817 &  0.752 &   0.779\\
Default + Strong & 0.795 &  0.831 & 0.781 & 0.793 \\
Default + Strong + Mixup & 0.842 & 0.829 &  0.790 & 0.795\\
Default + Strong + Mixup + TTA & \textbf{0.863}& \textbf{0.841} &  \textbf{0.815} & \textbf{0.821} \\
\bottomrule
\end{tabularx}
\label{tab:aug}
\end{table}
To select the best model for final LB submission, for each split, we created an ensemble of ResNet18 and MobileNetV3 with most effective augmentation as showin in Table. \ref{tab:aug} and finally ensembled predictions from all five splits as described in Section \ref{sec:method}.

\textbf{Comparison with top methods on final LB and conclusion \cite{stoic}}
The first ranked team pre-trained ConvNext  \cite{liu2022convnet} with MosMed \cite{morozov2020mosmeddata} for severity classification and UperNet \cite{xiao2018unified} with TCIA \cite{aerts2015data} for lesion segmentation. It was followed by training on STOIC dataset using metadata and the output of both backbones as vectors. They used a 5-fold cross-validation and ensemble model for testing. Team 2, used two vision encoders pre-trained on iBot  \cite{zhou2021ibot} via self-supervised learning on plain slices and segmented lung regions. They concatenated the features with age and sex features and used logistic regression for predictions. Team 3 used a lung segmentation model and autodidactic pre-training on segmented images.  The network's output was combined with age and passed to a fully connected layer and finally ensemble of five models and TTA was used.

In contrast to other methods, our approach did not use highly complex models with additional data sets. Despite its simplicity, our method is highly effective and competitive with the more complex techniques, as can be seen from Table \ref{tab:comaprison}.Unlike other methods, we did not use metadata for the final LB submission as it did not improve performance on the public data set. Despite observing the same phenomenon, other teams opted to include metadata in their final submission and it proved to be effective. If our approach had included metadata, it might have been ranked among the top three teams in the competition.

\begin{table}
\centering
\caption{
Final leader board results: comparison with top ranked methods \cite{stoic}}
\begin{tabular}{lcc}
\toprule
\multirow{2}{*}{\textbf{Model}} &\multicolumn{2}{|c}{\textbf{STOIC Public Data}} \\
 & \multicolumn{1}{|c|}{\textbf{AUC Severity}} & \multicolumn{1}{c}{\textbf{AUC COVID}} \\
 \midrule
First &   0.815 & 0.616 \\
Second & 0.811 &  0.845  \\
Third & 0.794 & 0.837 \\
\textbf{Fourth (our method) } &  0.787& 0.829 \\
\bottomrule
\end{tabular}
\label{tab:comaprison}
\end{table}

\section*{Acknowledgement}
This research was supported by Science Foundation Ireland under grant numbers 18/CRT/6183 (ML-LABS Centre for Research Training),18/CRT/6223, SFI/12/RC/2289{\_}P2 (Insight SFI Research Centre for Data Analytics),  13/RC/$2094\_P2$ (Lero SFI Centre for Software ) and 13/RC/$2106\_P2$ (ADAPT SFI Research Centre for  AI-Driven Digital Content Technology). For the purpose of Open Access, the author has applied a CC BY public copyright licence to any Author Accepted Manuscript version arising from this submission.

\bibliographystyle{apalike}

\bibliography{imvip}

\end{document}